\documentclass[a4paper]{article}
\usepackage{interspeech2011,amssymb,amsmath,epsfig}
\ninept
\def\reg{{\rm\ooalign{\hfil
     \raise.07ex\hbox{\scriptsize R}\hfil\crcr\mathhexbox20D}}}

\title{Joint Robust Voicing Detection and Pitch Estimation Based on Residual Harmonics}

\makeatletter
\def\name#1{\gdef\@name{#1\\}}
\makeatother
\name{{\em Thomas Drugman$^1$, Abeer Alwan$^2$}}

\address{$^1$TCTS Lab - FNRS researcher, University of Mons, Belgium \\
  $^2$Electrical Engineering Department, University of California, Los Angeles, United States of America\\
{\small \tt thomas.drugman@umons.ac.be}}

\begin{document}
\maketitle
\begin{abstract}
This paper focuses on the problem of pitch tracking in noisy conditions. A method using harmonic information in the residual signal is presented. The proposed criterion is used both for pitch estimation, as well as for determining the voicing segments of speech. In the experiments, the method is compared to six state-of-the-art pitch trackers on the Keele and CSTR databases. The proposed technique is shown to be particularly robust to additive noise, leading to a significant improvement in adverse conditions.
\end{abstract}

\section{Introduction}\label{sec:intro}

Pitch tracking refers to the task of estimating the contours of the fundamental frequency $F_0$ for voiced segments. Such a system is of particular interest in several applications of speech processing, such as speech coding, analysis, synthesis or recognition. While most current pitch trackers perform well in clean conditions, their performance rapidly degrades in noisy environments and the development of accurate and robust algorithms still remains a challeging open problem.


Techniques estimating $F_0$ from speech signals can be classified according to the features they rely on \cite{Furui}. Some methods use properties in the time domain, others focus on the periodicity of speech as manifested in the spectral domain, while a last category exploits both spaces. Besides, this information can be processed in a deterministic way, or using a statistical approach \cite{Furui}. This paper proposes a pitch tracking method exploiting the harmonics contained in the spectrum of the residual signal. The idea of using a summation of harmonics for detecting the fundamental frequency is not new. In \cite{Hermes}, Hermes proposed the use of a subharmonic summation so as to account for the phenomenon of virtual pitch. In \cite{Sun}, Sun suggested the use of the Subharmonic-to-Harmonic Ratio for estimating the pitch frequency and for voice quality analysis. The method proposed in this paper is different in several points. First, the spectrum of the residual signal (and not of the speech signal) is inspected. As in the Simplified Inverse Filter Tracking (SIFT) algorithm (which relies on the autocorrelation function computed on the residual signal, \cite{SIFT}), flattening the amplitude spectrum allows to minimize the effects of both the vocal tract resonances and of the noise. Secondly, the harmonic-based criterion used for the pitch estimation is different from those employed in the two aforementioned approaches. Besides the proposed criterion is also used for discriminating between voiced and unvoiced regions of speech. Note that harmonic-based Voice Activity Detection (VAD) has also been exploited in \cite{Lee}. 

The structure of the paper is the following. Section \ref{sec:Method} describes the principle of the proposed technique. An extensive quantitative assessment of its performance in comparison with other state-of-the-art techniques is given in Section \ref{sec:Exp}, focusing particularly on noise robustness. Section \ref{ssec:Protocol} presents the adopted experimental protocol. The implementation details of the proposed method are discussed in Section \ref{ssec:Optimization}. Methods compared in this work are presented in Section \ref{ssec:Methods} and results of the evaluation are provided in Section \ref{ssec:Results}.

\section{Pitch tracking based on residual harmonics}\label{sec:Method}

The proposed method relies on the analysis of the residual signal. Unlike SIFT \cite{SIFT} which is based on the autocorrelation function, this technique focuses on the residual harmonicity. Also, the proposed harmonic criterion is different from other comparable approaches, which makes the method efficient in adverse conditions both for voicing detection and pitch estimation.

For this, an auto-regressive modeling of the spectral envelope is first estimated from the speech signal $s(t)$ and the residual signal $e(t)$ is obtained by inverse filtering. This whitening process has the advantage of removing the main contributions of both the noise and the vocal tract resonances. For each Hanning-windowed frame, covering several cycles of the resulting residual signal $e(t)$, the amplitude spectrum $E(f)$ is computed. $E(f)$ has a relatively flat envelope and, for voiced segments of speech, presents peaks at the harmonics of the fundamental frequency $F_0$. From this spectrum, and for each frequency in the range $[F_{0,min},F_{0,max}]$, the Summation of Residual Harmonics (SRH) is computed as:

\begin{equation}\label{eq:SRH}
SRH(f)=E(f)+\sum_{k=2}^{N_{harm}} [{E(k\cdot f)-E((k-\frac{1}{2})\cdot f)}].
\end{equation}

Considering only the term $E(k\cdot f)$ in the summation, this equation takes the contribution of the first $N_{harm}$ harmonics into account. It could then be expected that this expression reaches a maximum for $f=F_0$. However, this is also true for the harmonics present in the range $[F_{0,min},F_{0,max}]$. For this reason, the substraction by $E((k-\frac{1}{2})\cdot f)$ allows to significantly reduce the relative importance of the maxima of $SRH$ at the even harmonics. The estimated pitch value $F_0^{*}$ for a given residual frame is thus the frequency maximizing $SRH(f)$ at that time.

\begin{figure}[!ht]
  \centering
  \includegraphics[width=0.48\textwidth]{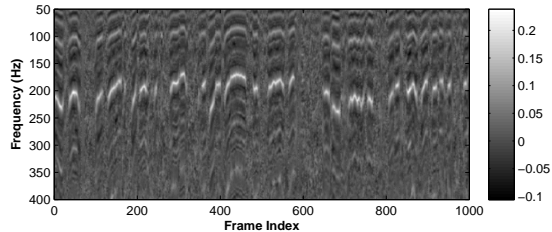}
  \caption{Evolution of $SRH$ for a segment of clean speech uttered by a female speaker.}
  \label{fig:SRH_example}    
\end{figure}

Figure \ref{fig:SRH_example} displays the typical evolution of $SRH$ for a segment of female voice. The pitch track (around 200 Hz) clearly emerges. Moreover, no particularly high value of $SRH$ is observed during the unvoiced regions of speech. Therefore, $SRH$ can also be used to provide voicing decisions by a simple local thresholding. More precisely, a frame is determined to be voiced if $SRH(F_0^{*})$ is greater than a fixed threshold $\theta$. Note that for the comparison with $\theta$, the residual spectrum $E(f)$ needs to be normalized in energy for each frame.

It is worth noting that, in Equation \ref{eq:SRH}, the risk of ambiguity with odd harmonics is not addressed. This may be problematic for low-pitched voices for which the third harmonic may be present in the initial range $[F_{0,min},F_{0,max}]$. Albeit we made several attempts to incorporate a correction in Equation \ref{eq:SRH} by substracting a term in $E((k\pm\frac{1}{3})\cdot f)$, no improvement was observed (this was especially true in noisy conditions). For this reason, the proposed algorithm works in two steps. In the first step, the described process is performed using the full range $[F_{0,min},F_{0,max}]$, from which the mean pitch frequency $F_{0,mean}$ of the considered speaker is estimated. In the second step, the final pitch tracking is obtained by applying the same process but in the range $[0.5\cdot F_{0,mean};2\cdot F_{0,mean}]$. It can be indeed assumed that a normal speaker will not exceed these limits. Note that this idea of restricting the range of $F_0$ for a given speaker is similar to what has been proposed in \cite{Yegna} (for the choice of the window length).

\begin{figure}[!ht]
  \centering
  \includegraphics[width=0.5\textwidth]{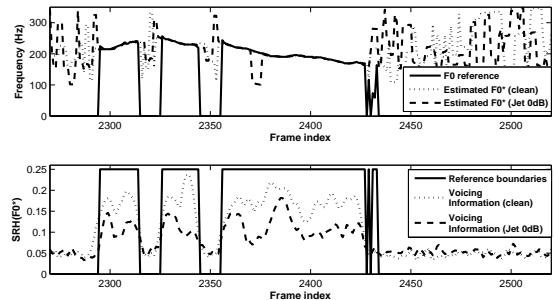}
  \caption{Illustration of the proposed method in clean and noisy speech (using a jet noise with a SNR of 0 dB). \emph{Top plot :} The pitch ground truth and the estimates $F_0^{*}$. \emph{Bottom plot :} The ideal voicing decisions and the values of $SRH(F_0^{*})$.}
  \label{fig:ExampleEvolution2}    
\end{figure}

Figure \ref{fig:ExampleEvolution2} illustrates the proposed method for a segment of female speech, both in clean conditions, and with a Jet noise at 0dB of Signal-to-Noise Ratio (SNR). In the top plot, the pitch ground truth and the estimated fundamental frequency $F_0^*$ are displayed. A close agreement between the estimates and the reference can be noticed during voiced speech. Interestingly, this is true for both clean and noisy speech (except on a short period of 5 frames where $F_0^*$ is half the actual fundamental frequency). It is worth noting that no post-correction of the pitch estimation, using for example dynamic programing, was applied. In the bottom plot, the values of $SRH(F_0^{*})$, together with the ideal voiced-unvoiced decisions, are exhibited since they are used for determining the voicing boundaries. It is observed that $SRH(F_0^{*})$ conveys a high amount of information about the voicing decisions. However, in adverse conditions, since the relative importance of harmonics becomes weaker with the presence of noise, the values of $SRH(F_0^{*})$ are smaller during voiced regions, making consequently the decisions more difficult.

\section{Experiments}\label{sec:Exp}

\subsection{Experimental Protocol}\label{ssec:Protocol}
The experimental protocol is divided into two steps: training and testing. The goal of the training phase is to optimize the several parameters used by the proposed algorithm described in Section \ref{sec:Method}. During the testing, the proposed method is compared to other state-of-the-art methods of pitch tracking, both in clean and noisy conditions. For assessing the performance of a given method, the four following measures are used \cite{Chu}:

The \textbf{Voicing Decision Error (VDE)} is the proportion of frames for which an error of the voicing decision is made.

The \textbf{Gross Pitch Error (GPE)} is the proportion of frames, where the decisions of both the pitch tracker and the ground truth are voiced, for which the relative error of $F_0$ is higher than a threshold of $20\%$.

The \textbf{Fine Pitch Error (FPE)} is defined as the standard deviation (in \%) of the distribution of the relative error of $F_0$ for which this error is below a threshold of $20\%$.

The \textbf{F0 Frame Error (FFE)} is the proportion of frames for which an error (either according to the GPE or the VDE criterion) is made. FFE can be seen as a single measure for assessing the overall performance of a pitch tracker.

%
%

The noisy conditions are simulated by adding to the original speech signal a noise at 0 dB of SNR. The noise signals were taken from the Noisex-92 database \cite{Noisex}. Since the main scope of this paper is the study of the robustness of pitch trackers, several types of noise were considered: speech babble, car interior, factory, jet cockpit, and white noise. 

During the \textbf{training} phase, the APLAWD database \cite{Lindsey1987} is used. It consists of ten repetitions of five phonetically balanced English sentences spoken by each of five male and five female talkers, with a total duration of about 20 minutes. The pitch ground truth was extracted by using the autocorrelation function on the parallel electroglottographic recordings. 

For the \textbf{testing}, both the Keele and CSTR databases were used, for comparison purpose with other studies. The Keele database \cite{Keele} contains speech from 10 speakers with five males and five females, with a bit more of 30 seconds per speaker. As for the CSTR
database \cite{CSTR}, it contains five minutes of speech from one male and one female speaker. For all datasets, recordings sampled at 16kHz were considered, and the provided pitch references were used as a ground truth.

\subsection{Parameter Optimization for the Proposed Method}\label{ssec:Optimization}
\begin{figure}[!ht]
  \centering
  \includegraphics[width=0.48\textwidth]{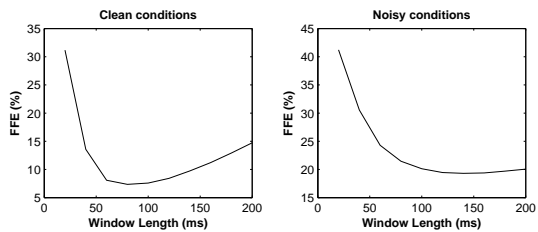}
  \caption{Influence of the window length on FFE, averaged in clean and noisy conditions.}
  \label{fig:WindowLength}    
\end{figure}

In this training step, each parameter is optimized so as to minimize the overall FFE, averaged over all speakers of the APLAWD database, and for both clean and noisy conditions. According to this objective framework, the optimal parameter values are the following. The LPC order for obtaining the residual signal by inverse filtering is set to 12, although it was observed not to have a critical impact in the range between 10 and 18. A too high order tends to overfit the spectral envelope, which may be detrimental in noisy conditions, while a too low value does not sufficiently remove the contributions of both the vocal tract and the noise. The optimal length for framing the residual signal is chosen to be 100 ms (while the frame shift is fixed to 10 ms). To illustrate this, Figure \ref{fig:WindowLength} shows the impact of the window length on the FFE for clean and noisy conditions. It turns out that a length of 100 ms makes a good compromise for being efficient in any environment. This means that our algorithm requires a large contextual information for performing well. Note that we observed that this does not affect the capabilities of the proposed method to track rapidly-varying pitch contours, maintaining low values of both GPE and FPE. The optimal number of harmonics used in Equation \ref{eq:SRH} is $N_{harm}=5$. Considering more harmonics is detrimental in adverse conditions, as the noise affects strongly the periodicity of the speech signal, and only the few first harmonic peaks emerge in the spectrum. Finally, the optimal threshold $\theta$ used for the voicing decisions is 0.07, as it gave the best tradeoff between false positive and false negative decisions.

\subsection{Methods compared in this work}\label{ssec:Methods}

In the following, the proposed technique (SRH) is compared to the seven following methods of pitch estimation and tracking:

\textbf{Get\_F0}: Included in the ESPS package, this method is an implementation of the RAPT algorithm \cite{Talkin}. In this work, we used the version available in Wavesurfer $<$http://www.speech.kth.se/wavesurfer/$>$.

\textbf{SHRP}: This spectral technique is based on the Subharmonic to Harmonic Ratio, as proposed in \cite{Sun}. For our tests, we used the implementation available in $<$http://mel.speech.nwu.edu/sunxj/pda.htm$>$.

\textbf{TEMPO}: This technique is based on a fixed point analysis \cite{Kawahara} and is available in the STRAIGHT toolkit $<$http://www. wakayama-u.ac.jp/\~{}kawahara/PSSws/$>$.

\textbf{AC}: This method relies on an accurate autocorrelation function and is implemented in the Praat toolbox $<$http://www.praat.org$>$. It was shown to outperform the original autocorrelation based and the cepstrum-based techniques \cite{AC}.

\textbf{CC}: This approach makes use of the crosscorrelation function \cite{CC} and is also implemented in the Praat toolbox.

\textbf{YIN}: This algorithm is one of the most popular and most efficient method of pitch estimation. It is based on the autocorrelation technique with several modifications that combine to prevent errors \cite{YIN}. Since YIN only provides F0 estimates, it is here \emph{coupled with the voiced-unvoiced decisions taken by our proposed SRH approach}. The YIN implementation can be freely found at $<$http://www.auditory.org/postings/2002/26.html$>$.

\textbf{SSH}: The \emph{Summation of Speech Harmonics} technique is given for comparison purpose as the proposed approach applied this time on the speech signal, and not on its residual as done in SRH. The contribution of the spectral envelope mainly due to the vocal tract is therefore not removed. Note that for SSH the optimal value of the threshold $\theta$ is 0.18.

All methods were used with their default parameter values for which they were optimized. The frame shift is fixed to 10 ms, and the range of F0 set to [50 Hz,400 Hz]. Note that we also made experiments with the AMDF and SIFT techniques, but these results are not included here due to space limitations, and since they provided among the worst performance in noisy environments.

\subsection{Results}\label{ssec:Results}

\begin{figure}[!ht]
  \centering
  \includegraphics[width=0.48\textwidth]{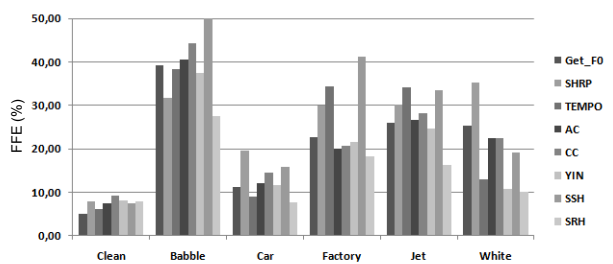}
  \caption{F0 Frame Error (\%) for \textbf{female speakers} and for all methods in six conditions: clean speech and noisy speech at 0dB of SNR with five types of noise.}
  \label{fig:Female}    
\end{figure}
\begin{figure}[!ht]
  \centering
  \includegraphics[width=0.48\textwidth]{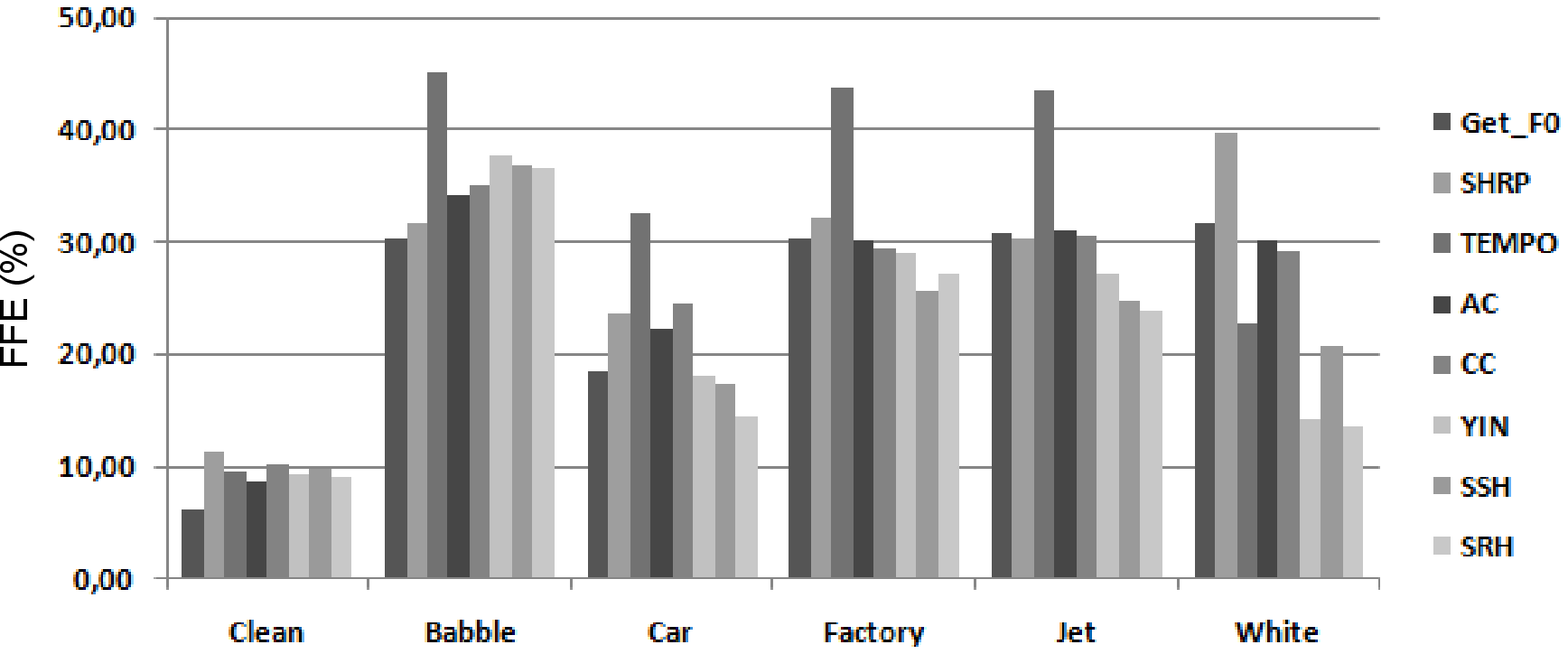}
  \caption{F0 Frame Error (\%) for \textbf{male speakers} and for all methods in six conditions: clean speech and noisy speech at 0dB of SNR with five types of noise.}
  \label{fig:Male}    
\end{figure}

Figures \ref{fig:Female} and \ref{fig:Male} show a comparison of the FFE (as it is an overall measure for assessing the performance of a pitch tracker) for all methods and in all conditions, respectively for female and male speakers. In clean speech, it is seen that the proposed SSH and SRH methods give a performance comparable to other techniques, while Get\_F0 outperforms all other approaches for both male and female speakers. On the opposite, the advantage of SRH is clearly noticed for adverse conditions. In 9 out of the 10 noisy cases (5 noise types and 2 genders), SRH provides better results than existing methods, showing generally an appreciable robustness improvement. The only unfavourable case is the estimation with a Babble noise for male speakers. This may be explained by the fact that this noise highly degrades the speech spectral contents at low frequencies. The five first residual harmonics used by SRH may then be strongly altered, leading to a degradation of performance. Inspecting the performance of SSH, it turns out that it exhibits among the worst results for female speakers in noisy environments, but is almost as efficient as SRH for male voices. This might be explained by the fact that for female voices, the first $N_{harm}$ harmonics cover a larger frequency width, on which the noise may have a more dramatic impact. This effect is reduced with SRH, since inverse filtering alleviates the noise influence.

\begin{table*}[!ht]
\centering
\begin{tabular}{| c || c | c | c | c || c | c | c | c ||| c | c | c | c || c | c | c | c |}
\hline
  & \multicolumn{8}{|c|||}{Clean conditions} & \multicolumn{8}{|c|}{Noisy conditions}\\
\hline
  & \multicolumn{4}{|c||}{Female} & \multicolumn{4}{|c|||}{Male} & \multicolumn{4}{|c||}{Female} & \multicolumn{4}{|c|}{Male}\\
\hline
  & VDE & GPE & FPE & FFE & VDE & GPE & FPE & FFE & VDE & GPE & FPE & FFE & VDE & GPE & FPE & FFE\\  
\hline
\hline
Get\_F0 & \textbf{3.74} & 2.78 & 2.95 & \textbf{4.92} & \textbf{5.34} & 1.79 & 3.06 & \textbf{6.11} & 20.8 & 14.8 & 2.4 & 24.9 & 27.7 & \textbf{2.7} & 2.7 & 28.3\\
\hline
SHRP & 7.01 & 2.03 & \textbf{2.52} & 7.83 & 10.2 & 2.74 & 3.17 & 11.4 & 27.0 & 11.5 & 1.9 & 29.3 & 30.1 & 6.8 & 2.8 & 31.5\\
\hline
TEMPO & 5.38 & 1.51 & 3.05 & 6.01 & 9.28 & \textbf{0.93} & 3.13 & 9.66 & 25.2 & 4.4 & 3.9 & 25.8 & 36.8 & 16.7 & 3.8 & 37.6\\
\hline
AC & 6.81 & 1.50 & 2.68 & 7.41 & 8.02 & 1.40 & \textbf{2.77} & 8.59 & 20.5 & 14.2 & 2.4 & 24.3 & 28.2 & 5.9 & \textbf{2.4} & 29.6\\
\hline
CC & 8.41 & 1.76 & 2.77 & 9.15 & 9.25 & 2.23 & 3.44 & 10.2 & 21.1 & 18.0 & 2.7 & 26.1 & 27.8 & 7.9 & 3.0 & 29.8\\
\hline
YIN & 7.29 & 1.88 & 2.95 & 8.06 & 8.34 & 2.47 & 2.93 & 9.38 & \textbf{15.1} & 19.0 & 3.0 & 21.2 & \textbf{22.1} & 11.9 & 2.8 & 25.2\\
\hline
SSH & 5.81 & 4.67 & 2.76 & 7.49 & 8.87 & 2.45 & 3.31 & 9.88 & 24.2 & 39.1 & \textbf{1.9} & 32.1 & 23.3 & 6.3 & 2.8 & 25.1\\
\hline
\textbf{SRH} & 7.29 & \textbf{1.29} & 3.10 & 7.81 & 8.34 & 1.95 & 3.46 & 9.15 & \textbf{15.1} & \textbf{2.7} & 2.6 & \textbf{16.0} & \textbf{22.1} & 4.0 & 2.7 & \textbf{23.1}\\
\hline
\end{tabular}
\caption{Detailed pitch tracking results in clean and noisy conditions (averaged over all noise types at 0 dB of SNR), for both male and female speakers.}
\label{tab:Complete}
\end{table*}

Table \ref{tab:Complete} presents the detailed results of pitch tracking for clean speech, and for noisy conditions (averaged over all noise types at 0dB of SNR). On clean recordings, Get\_F0 provides the best results in terms of VDE and FFE on both genders, while the best GPE is obtained by the proposed method SRH for female voices, and by TEMPO for male speakers. Regarding its efficiency in terms of FPE, albeit having the slightly largest values, SRH has a performance sensibly comparable to the state-of-the-art, confirming its ability to also capture the pitch contour details. On noisy speech, SRH clearly outperforms all other approaches, especially for female speakers where the FFE is reduced of at least 8.5\% (except for YIN which uses the proposed VAD from SRH). This gain is also substantial for male voices with regard to existing approaches (consequently leaving out of comparison the SSH and the modified YIN techniques), with a decrease of 5.3\% of FFE, and of 5.7\% regarding the errors on the voicing decisions. It is worth noting the remarkably good performance of SRH for female voices in noisy environments, providing very low values of VDE and GPE (and thus FFE). All methods (except SSH in adverse conditions) are also observed to give better results for female speakers than for male voices. Finally, it is interesting to emphasize that, while relying on the same voicing decisions, YIN leads in all conditions to a greater GPE than SRH, especially for noisy recordings. This confirms the quality of SRH both as a VAD and for pitch contour estimation.

\section{Conclusion}\label{sec:conclu}

This paper described a simple method of pitch tracking by focusing on the spectrum of the residual signal. A criterion based on the Summation of Residual Harmonics (SRH) is proposed both for pitch estimation and for the determination of voicing boundaries. A comparison with six state-of-the-art pitch trackers is performed in both clean and noisy conditions. A clear advantage of the proposed approach is its robustness to additive noise. In 9 out of the 10 noisy experiments, SRH is shown to lead to a significant improvement, while its performance is comparable to other techniques in clean conditions.


%
\eightpt
\bibliographystyle{IEEEtran}
\bibliography{strings,refs}

\end{document}